\documentclass[12pt,a4paper,DIV12]{scrartcl}
\usepackage[utf8]{inputenc}     
\usepackage[T1]{fontenc}        
\usepackage{lmodern}            
\usepackage[british]{babel}
\usepackage{amsmath}
\usepackage{amssymb}
\usepackage{amsfonts}
\usepackage{setspace}           
\usepackage{hyperref}
\usepackage{color}
\usepackage{authblk}
\title{%
  {\vspace{-20mm} \normalsize
   \hfill\parbox[b][30mm][t]{35mm}{\textmd{MS-TP-18-25}}}\\[-18mm]
The Leutwyler-Smilga relation on the lattice
\vspace*{3mm}}
\author[1]{Gernot M\"unster%
\thanks{munsteg@uni-muenster.de}}
\author[2]{Raimar Wulkenhaar%
\thanks{raimar@math.uni-muenster.de}}
\affil[1]{University of M\"unster, Institute for Theoretical Physics,
\authorcr
Wilhelm-Klemm-Str.~9, D-48149 M\"unster, Germany}
\affil[2]{University of M\"unster, Mathematical Institute,
\authorcr
Einsteinstr.~62, D-48149 M\"unster, Germany}
\date{July 5, 2019}
\newcommand{\I}{\ensuremath{\mathrm{i}}}
\newcommand{\E}{\ensuremath{\mathrm{e}}}
\newcommand{\Tr}{\ensuremath{\mathrm{Tr}}}
\newcommand{\1}{\ensuremath{\boldsymbol{1}}}
\newcommand{\arxiv}[2]{[arXiv:\,\href{http://arxiv.org/abs/#1}{\texttt{#1}} [\texttt{#2}]]}
\newcommand{\arxivold}[1]{[arXiv:\,\href{http://arxiv.org/abs/#1}{\texttt{#1}}\,]}

\begin{document}
\maketitle

\begin{abstract}
According to the Leutwyler-Smilga relation, in Quantum Chromodynamics (QCD)
the topological susceptibility vanishes linearly with the quark masses.
Calculations of the topological susceptibility in the context of lattice
QCD, extrapolated to zero quark masses, show a remnant non-zero value as a
lattice artefact. Employing the Atiyah-Singer theorem in the framework
of Symanzik's effective action and chiral perturbation theory, we show
the validity of the Leutwyler-Smilga relation in lattice QCD with
lattice artefacts of order $a^2$ in the lattice spacing $a$.
\end{abstract}
\vspace{5mm}
%
\section{Introduction}

Configurations of non-Abelian gauge fields in Euclidean space-time fall into
distinct topological sectors, characterised by the topological charge $Q \in
\mathbf{Z}$. The topological charge plays an important role in certain
phenomenological relations \cite{Witten:1979,Veneziano:1979,Vicari:2008jw}
and in the context of numerical simulations of gauge field theories on a
lattice, see e.\,g.~\cite{Muller-Preussker:2015daa}. A relevant observable
related to the topological charge is the topological susceptibility, defined
as $\chi_{t} = Q^2 /V$, where $V$ is a sufficiently large space-time volume.

In QCD with $N_f$ flavours of quarks, the topological susceptibility depends
on the light quark masses in a characteristic way. For simplicity let us
consider the case $N_f = 2$ with quark masses $m_u$ and $m_d$. Leutwyler and
Smilga \cite{Leutwyler:1992} showed that in the limit of small quark masses,
the topological susceptibility behaves as
\begin{equation}
\chi_{t} = \frac{m_u m_d}{m_u + m_d} \, \Sigma\,,
\end{equation}
where $\Sigma$ is a low-energy constant equal to the quark condensate. For
equal quark masses $m$, employing the Gell-Mann-Oakes-Renner relation, the
Leutwyler-Smilga relation reduces to
\begin{equation}
\chi_{t} = \frac{m}{N_f} \Sigma\ = \frac{1}{2 N_f} F_{\pi}^2 m_{\pi}^2
\end{equation}
with the pion mass $m_{\pi}$ and the pion decay constant $F_{\pi} \approx
93$ MeV.

The topological susceptibility and its dependence on the quark masses has
been studied by means of Monte Carlo simulations of lattice QCD in various
investigations, see e.\,g.~\cite{Bruno:2014ova,Alexandrou:2017bzk} and
references therein. Due to the finite lattice spacing $a$, lattice
corrections to the Leutwyler-Smilga relation appear. In particular,
extrapolation to the limit of vanishing quark masses leads to a remnant
value of $\chi_t$. For small quark masses the data can be fitted by
\begin{equation}
\chi_{t} = c_0 m_{\pi}^2 + c_1 a + c_2 a^2 \,,
\end{equation}
where $c_1 = 0$ for $O(a)$-improved Wilson fermions as e.\,g.\ in
\cite{Bruno:2014ova,Alexandrou:2017bzk}.
It is the purpose of this article to study $\chi_t$ in the framework
of Wilson chiral perturbation theory, and to derive the Leutwyler-Smilga 
relation including lattice artefacts. For simplicity we consider the
case of degenerate quark masses.

The Leutwyler-Smilga relation has its origin in the fact that, in accordance
with the Atiyah-Singer index theorem \cite{Atiyah:1968}, the massless
gauge-covariant Dirac operator $\gamma_{\mu} D_{\mu}$ in a background field
with topological charge $Q$ has 
a number of zero modes, such that $|Q|$ is the difference of the
numbers of left- and right-handed modes.
Let
\begin{equation}
D = \gamma_{\mu} D_{\mu}
+ P_L \mathcal{M} + P_R \mathcal{M}^{\dagger}
\end{equation}
be the massive Dirac operator, where $\mathcal{M} = m \1$ is the quark mass
matrix and $P_L = (\1 - \gamma_5)/2$ and $P_R = (\1 + \gamma_5)/2$ are the
chiral projectors. Then the partition function in a topological sector is
given by \cite{Leutwyler:1992}
\begin{align}
Z_Q &= \int \mathcal{D}A \ \E^{- S_G} \det D \\
&= \int \mathcal{D}A \ \E^{- S_G} \,\det{\!}'(D) \cdot
\begin{cases}
\det (\mathcal{M})^{Q} \,,& Q > 0\\
\det (\mathcal{M}^{\dagger})^{-Q} \,,& Q < 0,
\end{cases}
\end{align}
where $\det'(D)$ is the quark determinant without the mentioned zero modes.
The partition function in the $\theta$ vacuum,
\begin{equation}
Z(\theta) = \sum_Q \E^{\I\,\theta Q} Z_Q,
\end{equation}
can be obtained from $Z(0)$ by the replacement
\begin{equation}
\label{rule}
\mathcal{M} \rightarrow \mathcal{M}\, \E^{\,\I \theta / N_f},
\end{equation}
and $\chi_t$ is obtained from the vacuum energy $\epsilon(\theta) = - (1/V)
\ln Z(\theta)$ by
\begin{equation}
\label{epsilon-chi}
\chi_t = \left. 
\frac{\partial^2 \epsilon(\theta)}{\partial \theta^2} \right|_{\theta = 0}.
\end{equation}
Evaluation of $Z(\theta)$ by means of chiral perturbation theory in leading
order then leads to the Leutwyler-Smilga relation. Corrections in
next-to-leading order (NLO) have been calculated by Mao and Chiu
\cite{Mao:2009sy} and Guo and Mei{\ss}ner \cite{Guo:2015oxa}.

\section{Wilson chiral perturbation theory}

The effects of the lattice discretisation are taken into account in Wilson
chiral perturbation theory \cite{Rupak:2002sm}, for reviews see
\cite{Sharpe:2006pu,Golterman:2009kw}. The starting point is the Symanzik
effective theory \cite{Symanzik:1983dc} for lattice QCD. This effective
theory is a quantum field theory in the continuum. The lattice spacing
dependence of expectation values is encoded in a systematic expansion in
powers of the lattice spacing $a$. It has two origins. The first one is
Symanzik's effective Lagrangian, which is of the form
\begin{equation}
\mathcal{L}_{\text{eff}} = \mathcal{L}_{\text{QCD}} + a \mathcal{L}^{(1)}
+ a^2 \mathcal{L}^{(2)} + \ldots ,
\end{equation}
where $\mathcal{L}_{\text{eff}}$ is the continuum QCD Lagrangian, and the
$\mathcal{L}_k$ involve higher dimension operators. The second source are
lattice corrections to the measured operators, expanded in powers of $a$.
The leading lattice corrections to the QCD Lagrangian are given by adding
the term $a c \bar{\psi} \sigma_{\mu \nu} F^{\mu \nu} \psi$, with a certain
coefficient $c$, to the QCD Lagrangian in the continuum
\cite{Sheikholeslami:1985ij}. In the QCD Lagrangian the quark mass matrix
$\mathcal{M}$ is thus replaced by $\mathcal{M} + a c \sigma_{\mu \nu} F^{\mu
\nu}$. As the effective theory lives in the continuum, the topology of its
gauge fields is defined as in continuum QCD, and the Atiyah-Singer index
theorem holds. Therefore the partition function with vacuum angle $\theta$
is well defined. The matrices $\sigma_{\mu \nu}$ commute with $\gamma_5$,
and to first order in $a$ the partition function $Z(\theta)$ is obtained by
extending the substitution rule (\ref{rule}) to
\begin{equation}
\label{rule2}
\mathcal{M} \rightarrow \mathcal{M}\, \E^{\,\I \theta / 2}, \qquad
a \rightarrow a\, \E^{\,\I \theta / 2}.
\end{equation}
Concerning the observables, which are utilised for the numerical calculation
of $\chi_t$, we presume that they have lattice artefacts of order $a^2$ or
higher, which holds for the observables used in practice. Consequently, to
order $a$ the topological susceptibility is again given by the second
derivative (\ref{epsilon-chi}) of the vacuum energy in the effective theory.

The effective chiral Lagrangian, describing the physics of the pseudoscalar
mesons, is formulated in terms of the matrix-valued field $U(x) \in$ SU(2).
Its version for lattice QCD, defined in the continuum as well, is based on
Symanzik's effective theory. The quark masses and the lattice spacing $a$
enter the Lagrangian through the variables \cite{Rupak:2002sm}
\begin{equation}
\chi = 2 B_0 \mathcal{M}\,, \qquad
\qquad \rho = 2 W_0 a \1\,,
\end{equation}
where $B_0 = \Sigma / F_0^2$, $F_0$ is the pion decay constant to leading
order, and $W_0$ is an additional low-energy constant. In leading order (LO)
of chiral perturbation theory for lattice QCD, the Lagrangian is
\begin{equation}
\mathcal{L}_{2} 
= \frac{F_0^2}{4} \Tr(\partial_{\mu} U \, \partial_{\mu} U^\dagger)
- \frac{F_0^2}{4} \Tr(\chi U^\dagger + U \chi^\dagger)
- \frac{F_0^2}{4} \Tr(\rho \, U^\dagger + U \rho^\dagger).
\label{L2}
\end{equation}
For small quark masses, such that $m_{\pi}^{-1}$ is much larger than the
spatial extent of the system, the partition function is in LO dominated by
constant fields $U$ \cite{Leutwyler:1992},
\begin{equation}
Z = \int\! dU\ \E^{-V \mathcal{L}_{2}(U)}.
\end{equation}

Special unitary matrices can be diagonalised as
\begin{equation}
U=R^{\dagger} U_{\alpha} R,\quad
U_{\alpha} =
\begin{pmatrix}
\E^{\,\I \alpha} & 0 \\
0 & \E^{- \I \alpha}
\end{pmatrix},
\quad
R =
\begin{pmatrix}
\cos \frac{\beta}{2} & \sin \frac{\beta}{2} \\
-\sin \frac{\beta}{2} & \cos \frac{\beta}{2} 
\end{pmatrix}
\begin{pmatrix}
\E^{\,\I \gamma} & 0 \\
0 & \E^{- \I \gamma}
\end{pmatrix}.
\end{equation}
Taking the vacuum angle $\theta$ into account by means of the substitution
rule (\ref{rule2}), one obtains for degenerate quark masses
\begin{equation}
\mathcal{L}_{2}(U)=
\mathcal{L}_{2}(U_{\alpha}) =
- 2 F_0^2 (B_0 m + W_0 a) \cos \alpha \cdot \cos \frac{\theta}{2}.
\end{equation}
The group integral becomes
\begin{equation}
Z = \E^{2 V F_0^2 (B_0 m + W_0 a) \cos \frac{\theta}{2}}
\int_{-\pi}^{\pi} d\alpha\;j(\alpha)\ 
\E^{- 4 V F_0^2 (B_0 m + W_0 a) \sin^2 \frac{\alpha}{2} \cos \frac{\theta}{2}},
\end{equation}
where $j(\alpha)$ is the integral of the Jacobian of $d(R^{\dagger} U_\alpha
R)$ over $R$. The remaining integral can be estimated by $\text{const} (V
(m\Sigma + F_0^2 W_0 a))^{-1/2}$, giving
\begin{equation}
\epsilon (\theta) = - 2 F_0^2 (B_0 m + W_0 a) \cos \frac{\theta}{2}
+ O \left( \frac{\ln V}{V} \right).
\end{equation}
For large enough space-time volumes $V \gg 1/(m \Sigma)$, the first term
becomes exact (LO approximation, see \cite{Leutwyler:1992}). According to
Eq.~(\ref{epsilon-chi}) the resulting topological susceptibility is given by
\begin{equation}
\chi_t = \frac{F_0^2}{2} (B_0 m + W_0 a).
\end{equation}
Superficially this appears to imply lattice corrections of order $a$.
However, the same combination of low-energy constants enters the expression
for the pion mass in LO:
\begin{equation}
m_{\pi}^2 = 2 (B_0 m + W_0 a) + O(a^2).
\end{equation}
Therefore we obtain
\begin{equation}
\chi_t = \frac{F_0^2}{4} m_{\pi}^2 + O(a^2)
\end{equation}
in LO Wilson chiral perturbation theory. This confirms the coefficient of
the linear term to be the same as in continuum QCD, and shows order $a^2$
lattice artefacts.

The fact that the leading order $a$ terms in Wilson chiral perturbation theory
appear in the combination $B_0 m + W_0 a$ and can therefore be eliminated
in mesonic observables by a corresponding shift in the quark mass, can be seen
from the Lagrangian (\ref{L2}) and has first been observed by Sharpe and 
Singleton \cite{Sharpe:1998xm}.

Do the NLO contributions add terms of order $a$ to this expression? In NLO
there are one-loop contributions from $\mathcal{L}_{2}$ and tree-level
contributions from the next higher part $\mathcal{L}_{4}$ of the chiral
Lagrangian \cite{Gasser:1984}. Using the results from
\cite{Munster:2003ba,Scorzato:2004da,Sharpe:2004ny} it is easy to see that
the loop contributions do not produce $O(a)$ terms. Neglecting $O(a^2)$, the
pieces in $\mathcal{L}_{4}$ that could give rise to $O(a)$ tree-level
corrections, are
\begin{align}
\mathcal{L}_{4}^{a} =
&- W_6 \Tr(\chi U^{\dagger} + U \chi^{\dagger})\,
\Tr(\rho U^{\dagger} + U \rho^{\dagger}) \nonumber\\
&- W_7 \Tr(\chi U^{\dagger} - U \chi^{\dagger})\,
\Tr(\rho U^{\dagger} - U \rho^{\dagger}) \nonumber\\
&- W_8 \Tr(\chi U^{\dagger} \rho U^{\dagger} +
U \rho^{\dagger} U \chi^{\dagger}),
\end{align}
with certain low-energy constants $W_i$. Their contribution to $\chi_t$ is
\begin{equation}
16 (2 W_6 + 2 W_7 + W_8) B_0 m W_0 a 
= 8 (2 W_6 + 2 W_7 + W_8) W_0 m_{\pi}^2 a + O(a^2).
\end{equation}
The $O(a)$ expression vanishes in the limit $m_{\pi}^2 \to 0$, and
consequently for the remnant topological susceptibility in this limit we
conclude $c_1 = 0$.

\section{One-flavour QCD}

Leutwyler and Smilga \cite{Leutwyler:1992} also considered the case $N_f
=1$. In this case there is no spontaneously broken chiral symmetry and
chiral perturbation theory is not applicable. Their derivations are based on
different considerations of the vacuum energy. It is, however, alternatively
possible to study one-flavour QCD by means of partially quenched chiral
perturbation theory, as has been done in \cite{Farchioni:2007dw}. The theory
can artificially be extended by adding a number of quark flavours and the same
number of bosonic ghost quarks in such a way that the fermion determinant of
the additional quarks is cancelled by the functional integral over the ghost
quark fields. The simplest choice involves just one additional flavour, and
all quark masses $m$ taken to be equal. In the limit of vanishing quark
masses the theory possesses a graded
$\textrm{SU}(2|1)_L\,\otimes\,\textrm{SU}(2|1)_R$ symmetry, which is broken
spontaneously to a ``flavour'' supergroup SU(2$|$1). The pseudo-Goldstone
fields are parameterised by a graded matrix field $U(x)$ in SU(2$|$1).
Partially quenched chiral perturbation theory
can then be set up along the lines of
\cite{Sharpe:2001fh}, including O($a$) lattice effects. The effective action
for $U(x)$ is quite similar to the one of usual chiral perturbation theory,
the traces being replaced by supertraces. Without going into details, it
will be no surprise that in LO the relation
\begin{equation}
\chi_t = F_0^2 (B_0 m + W_0 a)
\end{equation}
holds. With the help of the divergence of the axial vector current and the
pseudoscalar density one can define a PCAC quark mass
$m_{\mathrm{\scriptscriptstyle PCAC}}$, which can be employed as a measure
of chiral symmetry breaking instead of the pion mass
\cite{Farchioni:2007dw}. Using $B_0 m_{\mathrm{\scriptscriptstyle PCAC}} =
B_0 m + W_0 a$ in LO, the $O(a)$ term is again cancelled in
\begin{equation}
\chi_t = F_0^2 B_0 m_{\mathrm{\scriptscriptstyle PCAC}} + O(a^2).
\end{equation}

\section{Conclusion}

In the context of lattice QCD with $N_f \geq 1$ flavours of quarks, we 
show in the framework of (partially
quenched) chiral perturbation theory that the topological susceptibility
$\chi_t$ obeys the Leutwyler-Smilga relation with the coefficient as in the
continuum, and that the lattice artefacts in the chiral limit appear to be
of order $a^2$.

\section*{Acknowledgements}

We thank Rainer Sommer and Gernot Akemann for discussions.


%
\end{document}